\documentclass[12pt, a4paper]{article}

\usepackage{epsfig}
\newcommand{\NP}[1]{{\it Nucl.\ Phys.}\ {\bf #1}}
\newcommand{\ZP}[1]{{\it Z.\ Phys.}\ {\bf #1}}
\newcommand{\PL}[1]{{\it Phys.\ Lett.}\ {\bf #1}}
\newcommand{\PR}[1]{{\it Phys.\ Rev.}\ {\bf #1}}

\newcommand{\IJMP}[1]{{\it Int.\ J.\ Mod.\ Phys.}\ {\bf #1}}
\newcommand{\be}{\begin{equation}}
\newcommand{\ee}{\end{equation}}
\newcommand{\bea}{\begin{eqnarray}}
\newcommand{\eea}{\end{eqnarray}}
\newcommand{\bfk}{\mbox{$\mathbf {k}$}}
\newcommand{\bfq}{\mbox{$\mathbf {q}$}}
\newcommand{\pup}{p^\uparrow}
\newcommand{\pdown}{p^\downarrow}
\newcommand{\qup}{q^\uparrow}
\newcommand{\qdown}{q^\downarrow}

\newcommand{\bfP}{\mbox{$\mathbf P$}} 
\newcommand{\Lup}{\Lambda^\uparrow} 
\newcommand{\Aup}{A^\uparrow} 
\newcommand{\hup}{h^\uparrow} 
\newcommand{\hdown}{h^\downarrow}

\begin{document}

\begin{center}
\bf{Cross sections and single spin asymmetries \\
in hadronic collisions with intrinsic \\
transverse momentum effects}\footnote{
Talk delivered by F. Murgia at the 
``IX Convegno su Problemi di Fisica Nucleare Teo\-ri\-ca'', 
October 9-12, 2002, Cortona, Italy.
}

\vspace*{0.6cm}

{\sf U.~D'Alesio and F.~Murgia}
\vskip 0.5cm

{\it Istituto Nazionale di Fisica Nucleare, 
Sezione di Cagliari \\
and Dipartimento di Fisica, Universit\`a di Cagliari \\
C.P. 170, I-09042 Monserrato (CA), Italy} \\
\end{center}

\vspace*{0.5cm}

\begin{abstract}
The role of intrinsic transverse momentum both in unpolarized and
polarized processes is discussed.
We consider inclusive cross sections for pion production in
hadronic collisions and for Drell-Yan processes; the results are
compared with available experimental data in several different kinematical
situations. 
We reanalyze transverse single spin asymmetries (SSA) 
observed in inclusive pion production, $\pup \, p \to \pi \, X$, in
terms of Sivers effect and 
show how it can be disentangled in polarized Drell-Yan processes  
by suitably integrating over some final configurations. 
Estimates for RHIC experiments are given.
\end{abstract}

\vspace*{0.5cm}

In the last years a lot of experimental and theoretical activity has
been devoted to the study of transverse single spin asymmetries (SSA)
in hadronic collisions and in semi-inclusive DIS.
In fact, perturbative QCD (pQCD) with ordinary
collinear partonic kinematics leads to negligible values
for these asymmetries, as soon as the relevant scale of the process under
consideration becomes large. There are however several experimental results
which seem to contradict this expectation. 
Among them let us mention: 
$i)$ the large polarization of $\Lambda$'s and other 
hyperons produced in $p \, N \to \Lup \, X$; 
$ii)$ the large asymmetry 
observed in $\pup \, p \to \pi \, X$ and  
$\bar p^\uparrow \, p \to \pi \, X$; 
$iii)$ 
the similar azimuthal asymmetry observed in  
$\ell \, \pup \to \ell \, \pi \, X$. 

A possible way out from this situation comes from
extending the collinear pQCD formalism with the inclusion of spin and
partonic intrinsic transverse momentum, $\bfk_\perp$, effects.
This leads to
the introduction of  new spin and $\bfk_\perp$ dependent
partonic distribution (PDF) and fragmentation (FF) functions, describing
fundamental properties of hadron structure \cite{anse}.

The role of $\bfk_\perp$ effects in inclusive hadronic reactions
has been extensively studied also in the calculation of
unpolarized cross sections. 
It has been shown that, particularly at moderately large $p_{_T}$ (which is the
region where SSA are measured to be large) these effects can be relevant and
may help in improving the agreement between experimental results and pQCD
(at LO and NLO) calculations, which often underestimate the data \cite{ww}. 

Based on these considerations, in this contribution we present a preliminary
account of an ongoing program which aims to describe consistently both
polarized and unpolarized cross sections (and SSA) for
inclusive particle production in hadronic collisions at
large energies and moderately large $p_{_T}$, using LO pQCD with
the inclusion of intrinsic transverse momentum effects.
Our main goal is not to fit the cross sections
but rather to show
that in our LO approach they are reproduced up to an overall factor
of 2-3, compatible with expected NLO K-factors and scale dependences,
which reasonably cancel out in SSA and are then out of our present interest.

In a LO pQCD approach at leading twist
with inclusion of 
$\bfk_\perp$ effects, the
unpolarized cross section for the inclusive process $A\,B\to C\,X$
reads 
\bea
d\sigma &\propto & \sum_{a,b,c,d} \hat f_{a/A}(x_a,\bfk_{\perp\,a}) \otimes
\hat f_{b/B}(x_b, \bfk_{\perp\,b}) \nonumber \\
&& \otimes\, d\hat\sigma^{ab \to c d}(x_a, x_b, \bfk_{\perp\,a},
 \bfk_{\perp\,b}) \otimes \hat D_{C/c}(z, \bfk_{\perp\,C})\,,
\label{ucr}
\eea
\noindent with obvious notations.
A similar expression holds for the numerator of a transverse
SSA ($\propto
d\,\Delta^{\!N}\!\sigma/d\sigma$), substituting for the polarized particle
involved the corresponding unpolarized PDF (or FF)
with the appropriate polarized one, $\Delta^{\!N}\!f$ or
$\Delta^{\!N}\!D$.
At leading twist there are four new spin and $\bfk_\perp$ dependent
functions to take into account:
\bea
\Delta^Nf_{q/\pup}(x, \bfk_{\perp})  &\equiv& 
\hat f_{q/\pup}(x, \bfk_{\perp})-\hat f_{q/\pdown}(x, \bfk_{\perp}) 
\>,
\label{delf1} \\
\Delta^Nf_{\qup/p}(x, \bfk_{\perp})   &\equiv& 
\hat f_{\qup/p}(x, \bfk_{\perp})-\hat f_{\qdown/p}(x, \bfk_{\perp})
\>, 
\label{delf2}\\
\Delta^N D_{h/\qup}(z, \bfk_{\perp})  &\equiv& 
\hat D_{h/\qup}(z, \bfk_{\perp}) - \hat D_{h/\qdown}(z, \bfk_{\perp}) 
\>,
\label{deld1} \\
\Delta^N D_{\hup/q}(z, \bfk_{\perp}) &\equiv& 
\hat D_{\hup/q}(z, \bfk_{\perp}) - \hat D_{\hdown/q}(z, \bfk_{\perp}) 
\>.
\label{deld2} 
\eea
The FF in Eq. (\ref{deld1}) is the Collins function \cite{col},
while the PDF in Eq. (\ref{delf1}) was first introduced by 
Sivers \cite{siv}.  
The function in Eq. (\ref{delf2}) was considered by Boer~\cite{boer} 
and the one in Eq. (\ref{deld2}) is 
the so-called ``polarizing'' FF \cite{abdm}.

Despite its successful phenomenology, the Sivers function was always
a matter of discussions and its very existence rather controversial; in fact
in Ref.~\cite{col} a proof of its vanishing was given, based on time-reversal 
invariance (notice that this does not apply to the FF sector). 
Ways out based on initial state interactions or 
non standard time-reversal properties \cite{dra} were discussed. 
Very recently a series of papers \cite{newsiv} 
have resurrected Sivers asymmetry 
in its full rights.  

The natural process to test the Sivers asymmetry is Drell-Yan where  
there cannot be any effect in fragmentation processes and, 
by suitably integrating over some final configurations, 
other possible effects vanish. 
SSA in Drell-Yan processes are particularly important now, as 
ongoing or imminent experiments at RHIC will be able to measure them. 

Let us start considering the role played by the intrinsic
$\bfk_{\perp}$ in the unpolarized cross sections. 
The PDF and FF in Eq. (\ref{ucr}) are given in a
simple factorized form, and the $\bfk_\perp$ dependent part
is usually taken to have a Gaussian shape:
\begin{equation}
\hat f_{a/A}(x,\bfk_{\perp a}) = f_{a/A}(x)\,\frac{\beta^2}{\pi}\>
e^{-\beta^2\,k_{\perp a}^{\,2}}\>;
\>\>
\hat D_q^h(z,\bfk_{\perp h}) = D_q^h(z)\,\frac{\beta'^2}{\pi}\>
e^{-\beta'^2\,k_{\perp h}^{\,2}}\,,
\label{gk}
\end{equation}
\noindent where the parameter $\beta$ ($\beta'$) is related to the
average partonic (hadronic) $k_{\perp}$ by the simple relation  
$1/\beta(\beta')=\langle\,k_{\perp a(h)}^2\,\rangle^{1/2}$.
Similar expressions are adopted for the polarized PDF and FF of
Eq.s (\ref{delf1})--(\ref{deld2}).

We have considered 
several hadronic processes, analyzing a large sample of available data
in different kinematical situations.
Here we limit ourselves to present few indicative results and comments
regarding: 
{\it i)} The Drell-Yan process $p\,p\to \ell^+\ell^-\,X$;  
{\it ii)} Inclusive pion production in $p\,p\to\pi\,X$.\footnote{A 
full account of this analysis, including
prompt photon production in $p\,p\to\gamma\,X$, will be presented 
elsewhere \cite{damu}.}

We find that an overall good reproduction of the corresponding
unpolarized cross sections is possible (within the limits indicated
above) by choosing, depending on the kinematical situation
considered,
$\beta=1.0-1.25$ (GeV/$c)^{-1}$ (that is,
$\langle\,k_\perp^2\,\rangle^{1/2}=0.8-1.0$ GeV/$c$).
The choice of $\beta$ is related to the set
of PDF utilized; throughout this paper we use the
GRV94 set \cite{grv94}. The optimal choice of $\beta'$ in case
{\it ii)} (pion production) is commented below. 

\noindent {\it i)}  At LO and within collinear partonic configuration 
the final lepton pair produced in Drell-Yan processes
cannot have any transverse momentum, $q_{_T}$, with respect to the
colliding beams. Experimental data show however that the lepton pair
has a well defined $q_{_T}$ spectrum.
As an example, in Fig. 1a we show estimates of the invariant cross section
at $E$ = 400 GeV as a function of $q_{_T}$,
for several different invariant mass bins (in GeV) at fixed rapidity
$y$ = 0.03, and using $\beta=1.11$ (GeV$/c)^{-1}$;
data are from \cite{ito}.  
Theoretical curves are arbitrarily raised by a factor $K_{\rm fac}=1.6$,
which could be well accommodated by NLO K-factors and scale dependences,
an issue that as said above we do not address here.
Notice how data are well reproduced by a
Gaussian dependence up to $q_{_T} =2-2.5$ GeV/$c$; larger $q_{_T}$ data
show a power-law decrease well explained by pQCD corrections. 

\begin{figure}[ht]
\begin{center}
\mbox{~\psfig{file=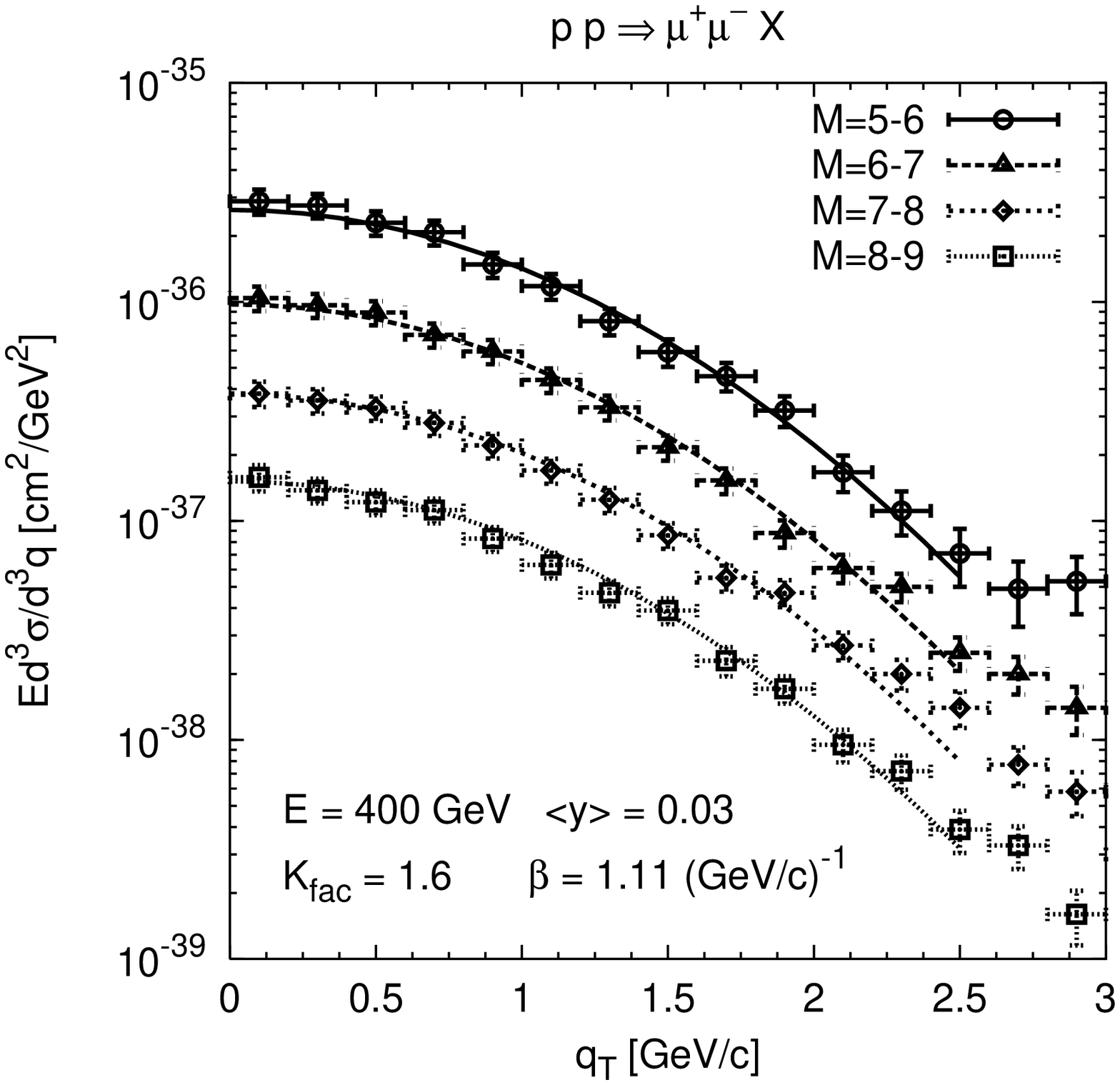,angle=-90,width=.5\textwidth}
\psfig{file=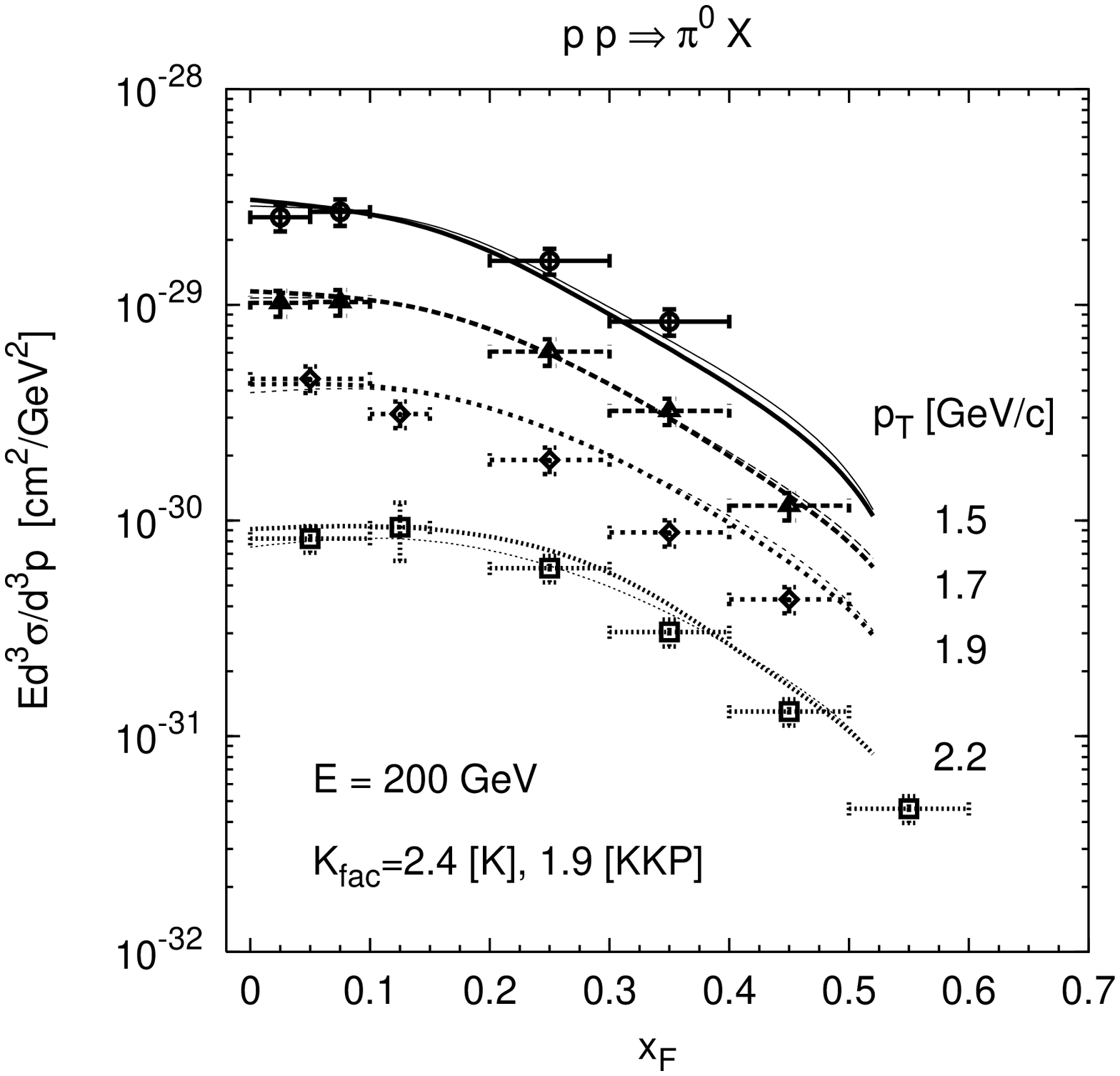,angle=-90,width=.5\textwidth}}
\caption{\small The invariant cross section for (a) $p\,p\to\mu^+\mu^-\,X$
vs. $q_{_T}$ and (b) $p\,p\to\pi^0\,X$ vs. $x_{_F}$; see plots and text for
more details. 
}
\end{center}
\end{figure}

\noindent {\it ii)}
For inclusive pion production, $p\,p\to\pi\,X$,
some experimental results for SSA are also available, and we
can see how our approach works for SSA and unpolarized cross sections
at the same time. This case is however more intricate, since we can
have $\bfk_\perp$ effects in the fragmentation process also.
The $z$ and $\bfk_\perp$ dependences in the FF are chosen according
to Eq. (\ref{gk}); a direct $z$ dependence of the $\beta'$ parameter
seems to be favored,
$1/\beta'(z)=\langle\,k_{\perp \pi}^{\,2}(z)\,\rangle^{1/2}=
1.4\,z^{1.3}\,(1-z)^{0.2}$ GeV/$c$.

Unpolarized FF are presently known with much less accuracy than
nucleon PDF. In particular, all available sets of parameterizations 
for the pion FF are for neutral pions
(or for the sum of charged pions), since $e^+e^-$ data do not allow
to separate among $\pi^+$ and $\pi^-$;
this can be made under further assumptions, which remain to be tested.
In Fig. 1b we present estimates of the invariant
cross section for the process $p\,p\to\pi^0\,X$ at $E$= 200 GeV 
vs. $x_{_F}$ for different $p_{_T}$ values.
We use two sets of FF from Kretzer (K, thin lines) \cite{kre} and
 Kniehl, Kramer, and P\"otter (KKP, thick lines) \cite{kkp},  
$K_{\rm fac}=2.4$(K), 1.9(KKP), $\beta=1.25$ (GeV$/c)^{-1}$.
Data are from \cite{dona}.

Let us now consider the SSA in $p^\uparrow\,p\to\pi\,X$, within the same
approach and assuming it is generated by the Sivers effect alone,
that is from a spin and $\bfk_\perp$ effect in the PDF
inside the initial polarized proton, described by the Sivers function
$\Delta^{\!N}\!f_{q/p^\uparrow}(x,\bfk_\perp)$. Other possible
sources for SSA, like the so-called Collins effect, concerning
the fragmentation of a polarized parton into the final observed pion, 
are not considered here.
Analogous studies have already been performed \cite{abm}, 
using an effective averaging on
$\bfk_\perp$ and a simplified partonic kinematics. Here we show
the first results with full treatment of $\bfk_\perp$ effects
and partonic kinematics. These results are in good qualitative
agreement with previous work. 

The numerator of the SSA, $d\sigma^\uparrow-d\sigma^\downarrow$ can be
expressed in the form of Eq.~(\ref{ucr}), with the substitution
$\hat f_{a/A}(x,\bfk_\perp)\,\to\,
\Delta^{\!N}\!f_{q/p^\uparrow}(x,\bfk_\perp)$. 
For the Sivers function we choose an expression similar to that of the
unpolarized distribution, Eq.~(\ref{gk})
\begin{equation}
\Delta^{\!N}\!f_{q/p^\uparrow}(x,\bfk_\perp)=
\Delta^{\!N}\!f_{q/p^\uparrow}(x)\,h(k_\perp)\,\sin\phi_{k_\perp}\>,
\label{dfh}
\end{equation}
\noindent where $\phi_{k_\perp}\!$ is the angle between $\bfk_\perp$ and the
polarization vector of the proton; $\Delta^{\!N}\!f_{q/p^\uparrow}(x)$
and $h(k_\perp)$ are such to fulfill the general 
positivity bound 
$|\Delta^{\!N}\!f_{q/p^\uparrow}(x,k_\perp)|/
2\,\hat f_{q/p}(x,k_\perp)\leq 1$:
\begin{equation}
\Delta^{\!N}\!f_{q/p^\uparrow}(x) = N_q\,x^{a_q}(1-x)^{b_q}\,
\frac{(a_q+b_q)^{(a_q+b_q)}}{a_q^{a_q}\,b_q^{b_q}}\,
2\,f_{q/p}(x)\,,\quad
|N_q|\leq 1\,
\label{fx}
\end{equation}
\begin{equation}
h(k_\perp)=\left(2\,e\,\frac{1-r}
{r}\right)^{1/2}\,\frac{\beta^3}{\pi}\,
k_\perp\,\exp\left[\,-\beta^2k_\perp^2/r\,\right]\,
, \quad 0 < r  < 1\>.
\label{hk}
\end{equation}

A choice of the parameters in Eq.s (\ref{fx}),(\ref{hk})
which allow to reasonably reproduce the experimental results
for the pion SSA is the following (only valence quark contributions
to the Sivers function are considered):
\begin{eqnarray}
N_u &=& +0.5\quad a_u=2.0\quad b_u=0.3 \label{nab}\\
N_d &=& -1.0\quad a_d=1.5\quad b_d=0.2\,,
\quad\quad\quad r \simeq 0.7\>.\nonumber
\end{eqnarray}

In Fig. 2a we show our preliminary estimates of $A_N$ with Sivers effect
at $E$ = 200 GeV and $p_{_T}$= 1.5 GeV/c, vs. $x_{_F}$, 
for three different choices of the pion FF:
K, KKP and a modified version of K. Data are from \cite{e704}.
The SSA for $\pi^+$ and $\pi^0$ is well reproduced independently of the
FF set. 
Interestingly, the $\pi^-$ case shows a stronger
sensitivity to the relation between the leading and non-leading contributions
to the fragmentation process, which cannot be extracted from present
experimental information. 
In fact, our results with the K(KKP) FF sets underestimate
(overestimate) in magnitude the
$\pi^-$ asymmetry, while a good agreement is recovered using a somehow
fictitious set (K-mod) with an intermediate behavior.

A more direct way to extract the Sivers asymmetry 
is the analysis of SSA in 
Drell-Yan processes, that is the production of 
$\ell^+\ell^-$ pairs in the collision of two hadrons $A$ 
and $B$ \cite{adm02}.    
By considering differential cross sections 
in the squared invariant mass ($ M^2 = (p_a + p_b)^2 $), 
rapidity ($ y $) and  
transverse momentum of the lepton pair ($ \bfq_{_T} $) 
and integrating out the di-lepton angular dependence, 
the difference between 
$d\sigma^\uparrow$ for $A^\uparrow \, B \to \ell^+ \, \ell^- \, X$ 
and $d\sigma^\downarrow$ for $A^\downarrow \, B \to \ell^+ \, \ell^- \, X$ 
from the Sivers asymmetry of Eq. (\ref{delf1}), 
is
\bea
&& d\sigma^\uparrow  -  d\sigma^\downarrow  =  \\
&& \sum_{ab} \int \left[ dx_a \, d^2\bfk_{\perp a} 
\, dx_b \, d^2\bfk_{\perp b} \right] \, 
\Delta^Nf_{a/A^\uparrow}(x_a,\bfk_{\perp a}) \,
\hat f_{b/B}(x_b,\bfk_{\perp b}) \,
d\hat\sigma^{ab \to \ell^+\ell^-}\nonumber
\!.
\label{ddy1}
\eea
We take the hadron $A$ as moving along the positive $z$-axis, 
in the $A$-$B$ c.m. frame and the transverse polarization 
of hadron $A$, $\bfP_{\!A}$, along the $y$-axis. 

\begin{figure}[t]
\begin{center}
\hspace*{.2cm}
\mbox{~\psfig{file=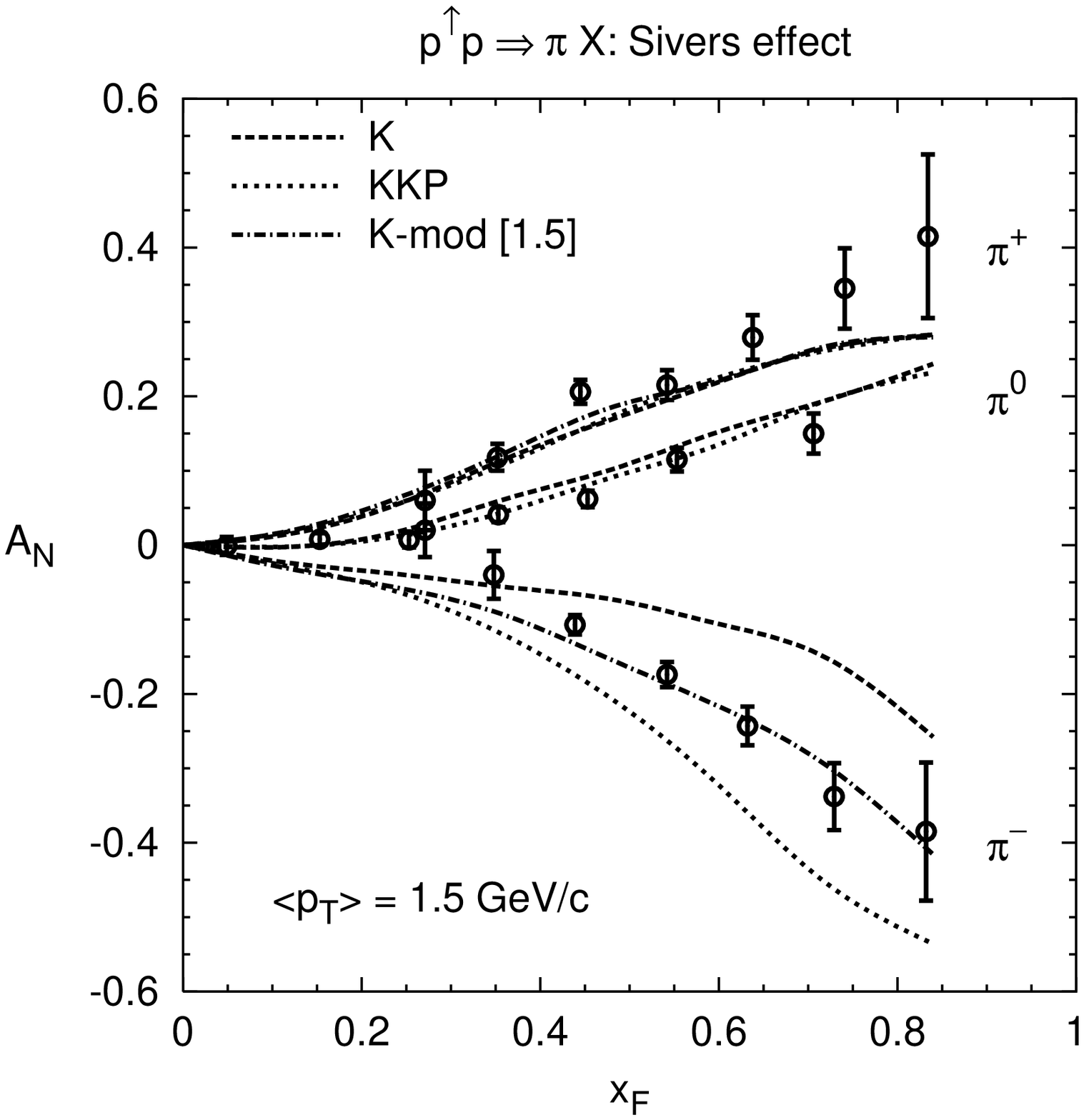,angle=-90,width=.5\textwidth}
\psfig{file=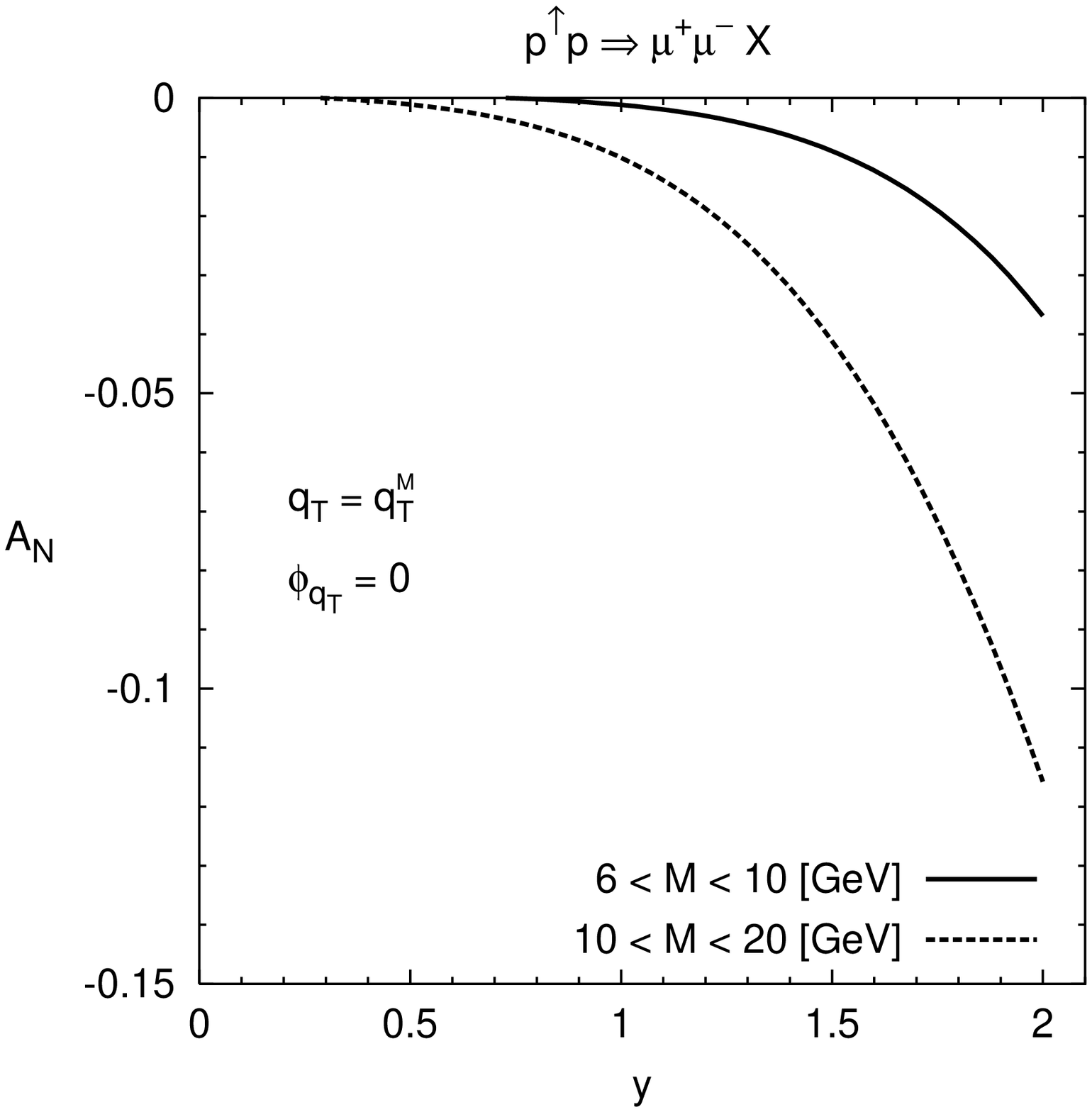,angle=-90,width=.5\textwidth}}
\caption{\small The SSA for (a) $p^\uparrow\,p\to\pi\,X$ vs. $x_{_F}$ 
and (b) $p^\uparrow\,p\to\mu^+\mu^-\,X$ vs. $y$; see plots and text for more
details. 
}
\end{center}
\end{figure}

In the kinematical regions such that: 
$q_{_T}^2 \ll M^2 \ll M_Z^2$ and $k_{\perp a,b}^2 \simeq q_{_T}^2$, 
the asymmetry becomes
\be
A_N = \frac
{\sum_q e_q^2 \int d^2\bfk_{\perp q} \, d^2\bfk_{\perp \bar q} \>
\,\delta^2_{k_{_\perp}\! q_{_T}}\,
\Delta^Nf_{q/\Aup}(x_q, \bfk_{\perp q}) \>
\hat f_{\bar q/B}(x_{\bar q}, \bfk_{\perp \bar q})}
{2 \sum_q e_q^2 \int d^2\bfk_{\perp q} \, d^2\bfk_{\perp \bar q} \>
\,\delta^2_{k_{_\perp}\! q_{_T}}\,
\hat f_{q/A}(x_q, \bfk_{\perp q}) \>
\hat f_{\bar q/B}(x_{\bar q}, \bfk_{\perp \bar q})} \>, \label{ann}
\ee
where $\delta^2_{k_{_\perp}\! q_{_T}} \equiv  
\delta^2(\bfk_{\perp q} + \bfk_{\perp \bar q} - \bfq_{_T})$, 
$x_q\simeq\frac{M}{\sqrt s} \, e^y $ 
and $x_{\bar q}\simeq\frac{M}{\sqrt s} \, e^{-y}$, with $a,b = q, \bar q$  
and $q = u, \bar u, d, \bar d, s, \bar s$.    

Notice that other sources of SSA, like i.e. 
the distribution function in Eq. (\ref{delf2}),   
would lead to a contribution to $A_N$ 
that vanishes upon integration 
over all final angular configurations of the $\ell^+\ell^-$ pair.

Inserting the above choice of $\Delta^Nf(x, \bfk_\perp)$ and 
$\hat f(x, \bfk_\perp)$ 
into Eq. (\ref{ann}) one can perform analytical
integrations; being $\beta$ independent of $x$ one gets 
\bea
A_N(M,y,\bfq_{_T}) &=&
{\mathcal Q}(q_{_T},\phi_{q_{_T}})\>{\mathcal A}(M,y) \nonumber\\
&=& \frac{2r \sqrt{2\,e\,r\,(1-r)}}{(1+r)^2}\, 
\beta \,q_{_T} \,\cos\phi_{q_{_T}}\>\exp\,\left[\,-\frac{1}{2}\,\frac{1-r}{1+r}\,
\beta^2\,q_{_T}^2\,\right]\nonumber\\
&\times&\>\frac{1}{2}\>\frac{\sum_q e_q^2 \, \Delta^Nf_{q/\pup}(x_q) \,
f_{\bar q/p}(x_{\bar q})} {\sum_q e_q^2 \, f_{q/p}(x_q) \, 
f_{\bar q/p}(x_{\bar q})} \> ,
\label{anr}
\eea
where $\phi_{q_{_T}}$ is the azimuthal angle of $\bfq_{_T}$. 
${\mathcal Q}(q_{_T})$ has a maximum at 
$q_{_T} = q_{_T}^M = \sqrt{(1+r)/(1-r)}/\beta $,  
where its value is ${\mathcal Q}(q_{_T}^M)\equiv {\mathcal Q}_M = 
[\,2\,r/(1+r)\,]^{3/2}$. Notice that only the position of the maximum
depends on  $\beta$.

One further uncertainty concerns the sign of the asymmetry: as noticed 
by Collins and verified by Brodsky \cite{newsiv}, the Sivers asymmetry 
has opposite signs in Drell-Yan and SIDIS, respectively related 
to $s$-channel and $t$-channel elementary reactions. As in $p-p$ 
interactions we expect that large $x_{_F}$ regions are dominated by 
$t$-channel quark processes, we think that the Sivers function
extracted from $p-p$ data should be opposite to that contributing
to D-Y processes. Our numerical estimates will then be given with the same
parameters as in Eq.~(\ref{nab}), {\it changing the signs of} $N_u$ and $N_d$. 
Given these considerations, even a simple comparison of the sign of our
estimates with data might be significant.

In Fig. 2b we show $A_N$ at $\sqrt s = 200$ GeV 
as a function of 
$y$ averaged over two kinematical ranges $6\leq M \leq 10$ GeV 
and $10 \leq M \leq 20$ GeV. 
We have fixed $q_{_T}=q_{_T}^M$ ($\simeq 1.9$ GeV/$c$), and 
$\phi_{q_{_T}}=0$, which maximizes the $\bfq_{_T}$-dependent part of the 
asymmetry; on the other hand $A_N$ is reduced by a factor of 50\%  
at $q_{_T}\simeq 0.6$ GeV/$c$.

We can also consider the asymmetry averaged
over $\bfq_{_T}$ up to a value of $q_{_T} = q_{_{T1}}$ 
(integrating over $\phi_{q_{_T}}$ in the range $[0,\pi/2]$ only,
otherwise one would get zero). 
In our simple model 
for $q_{_{T1}} \ge 1.7 $ GeV/$c$ we would get $\langle A_N \rangle\simeq 0.4
\,A_N(q_{_T}^M)$ (for $q_{_{T1}} = 0.6 $ GeV/$c$ $\langle A_N \rangle\simeq 0.2
\,A_N(q_{_T}^M)$ ).

Our estimates show that $A_N$ can be well measurable within 
RHIC expected statistical accuracy; the actual values depending on the
assumed functional form of the Sivers function and its role with 
valence~quarks.

In conclusion, we have presented here preliminary results of 
a study of partonic transverse momentum effects
both in unpolarized and polarized cross sections (and SSA) for inclusive
particle production in hadronic collisions. 
These results show that it seems possible to reproduce 
reasonably well, within pQCD at LO and leading twist 
and up to a factor of 2-3,
unpolarized cross sections for Drell-Yan processes, 
and inclusive pion production in hadronic collisions, 
in several different kinematical situations. 
Within the same approach, we have reanalyzed the 
SSA for $p^\uparrow\,p\to\pi\,X$ taking into account 
Sivers effect alone and
found reasonable agreement with data and 
with previous (simplified) theoretical results 
that are therefore confirmed by our analysis.
We have also shown how SSA in Drell-Yan processes can be a powerful
tool to learn on the Sivers asymmetry.

The next steps of this program are the study of the pion SSA with
Collins effect, of the SSA in photon production,
and of the unpolarized cross section and the
transverse polarization for $\Lambda$ production in
unpolarized hadronic collisions.
The extension of our analysis to RHIC kinematics,
where a thorough program on SSA measurements is in progress, 
is of great interest. First estimates of the SSA in
$p^\uparrow\,p\to\pi\,X$ seem to be in reasonable agreement with
preliminary results from RHIC \cite{rak}.

\section*{\small Acknowledgments}

\vspace*{-5pt}

This contribution is based on work done in collaboration with M. Anselmino.
Support from COFIN. MURST-PRIN is acknowledged.


{\small
\begin{thebibliography}{0}
%
\bibitem{anse}  For review papers on the subject, see, e.g.,
                Z.-T. Liang and C. Boros, \IJMP
                {A15}, 92 (2000); M. Anselmino, e-Print
                Archive: hep-ph/0201150. 

\vspace*{-7pt}

\bibitem{ww}    X.-N. Wang, \PR {C61}, 064910 (2000);
                C.-Y. Wong and H. Wang, \PR {C58}, 376 (1998);
                Y. Zhang, G. Fai, G. Papp, G. Barnaf\"oldi and
                P. L\'evai, \PR {C65}, 034903 (2002);
\vspace*{-7pt}

\bibitem{col}   J.C. Collins, \NP {B396}, 16 (1993).

\vspace*{-7pt}

\bibitem{siv}   D. Sivers, \PR {D41}, 83 (1990); \PR {D43}, 261 (1991).

\vspace*{-7pt}

\bibitem{boer}  D. Boer, \PR {D60}, 014012 (1999).

\vspace*{-7pt}

\bibitem{abdm}  M. Anselmino, D. Boer, U. D'Alesio and F. Murgia,
                \PR {D63}, 054029 (2001); \PR {D65}, 114014 (2002).

\vspace*{-7pt}

\bibitem{dra}   M. Anselmino, V. Barone, A. Drago and F. Murgia,  
		e-Print Archive: hep-ph/0209073.   
\vspace*{-7pt}

\bibitem{newsiv}  S.J. Brodsky, D.S. Hwang and I. Schmidt,  
		  \PL {B530}, 99 (2002); \NP {B642}, 344 (2002);  
		  J.C. Collins, \PL {B536}, 43 (2002);
		  X. Ji and F. Yuan,  \PL {B543} 66, (2002).

\vspace*{-7pt}

\bibitem{grv94} M. Gluck, E. Reya and A. Vogt, \ZP {C67}, 433 (1995).

\vspace*{-7pt}

\bibitem{damu}  U. D'Alesio and F. Murgia, work in preparation.

\vspace*{-7pt}

\bibitem{ito}   A.S. Ito {\em et al.}, \PR {D23}, 604 (1981).

\vspace*{-7pt}

\bibitem{kre}   S. Kretzer, \PR {D62}, 054001 (2000).

\vspace*{-7pt}

\bibitem{kkp}   B.A. Kniehl, G. Kramer and B. P\"{o}tter,
                \NP {B582}, 514 (2000).

\vspace*{-7pt}

\bibitem{dona}  G. Donaldson {\em et al.}, \PL {B73}, 375 (1978).

\vspace*{-7pt}

\bibitem{abm}   M. Anselmino, M. Boglione and F. Murgia, 
                \PL {B362}, 164 (1995); M. Anselmino and F. Murgia, 
                \PL {B442}, 470 (1998).

\vspace*{-7pt}

\bibitem{e704}  D.L. Adams {\em et al.} (E704 Collab.),
                \PL {B261}, 197 (1991); \PL {B264}, 462 (1991).

\vspace*{-7pt}

\bibitem{adm02}  M. Anselmino, U. D'Alesio and F. Murgia, 
		 e-Print Archive: hep-ph/0210371.

\vspace*{-7pt}

\bibitem{rak}   G. Rakness, in proceedings of the 
		15th International Spin Physics Symposium (SPIN 2002), 
		Long Island, New York, 9-14 Sep. 2002; L.C. Bland, 
                {\it ibidem}.
%
\end{thebibliography}
}
\end{document}